# Translation of Bengali Terms in Mobile Phones: a Simplified Approach Based on the Prescriptions of Conventional Accent Understand Ability

Partha Pratim Ray

*CSE Department, Surendra Institute of Engineering and Management*
*Siliguri, West Bengal, India*

parthapratimray@hotmail.com

*Abstract*— **Technology is the making, usage and knowledge of tools, techniques, crafts, systems or methods of organization in order to solve a problem or serve some purpose. This is true for humanitarian issues also. Such as the issue of language and its primitive attraction for its native speakers which is visible in the cases of the language spoken at home, outside home, in its choice of newspapers, and TV channels. Everyone finds to accomplish its need by the same way. Example includes the preference of using mobile phones in English. The satisfactory answer to this tendency may be the lack of finding the translations in native language---Bengali terms used in current mobile phones are hard to understand by users. I have investigated various mobile phone models available in Indian market which have lot of problems in Bengali interpretation. I have sort out the root cause of this problem to be the conventional accent understand ability. Depending on this I have created a set of equivalent terms that I hope to be simpler in use. In this paper I have performed experiments to compare the new terms to the available ones. Our findings show that the newly derived terms do better in term of performance than to current ones. It has also been seen that acceptance of Bengali terms in mobile phones might grow if the parameter of simpler and conventional accent understand ability are met while designing.**

## I. INTRODUCTION

Bengali is the second most spoken language in India, next to Hindi. Almost 83,369,769 native people use this language to do day to day work from speaking to voice communicating to chatting. This huge number is 8.11% of total population of India, as per census 2001. 250 million people of Bangladesh, India, United Arab Emirates and Saudi Arabia speak in Bengali [1]. This craze to Bengali has positioned in the 6th most spoken native language in the world [2]. As per the census 2011, 26% people of India are illiterate and about 20.43 people who talk in Bengali, have not even signed their name for first time [3]. Only 0.02% of Indian people speak in English as first language; however they are not always the users in term of socio-economical aspect. From here it is very clear that English is not most preferable language in Indian scenario.

Due to fantastic advent of mobile cellular service, the overall teledensity at the end of March 2010 was 52.74%, which accompanied with large wireless subscriber base about 58.43% of total Indian population [4]. The current telecom growth has come in form of voice communication, though the other adaptive ways for communication technology lie in short message service (SMS), voice mail, chatting, blogging etc. totally comprised in English language [5]. One such example is that a regional daily soap program in national TV channel asks for feedback in form of English SMS. This leads to design for easy translation scheme for mobile phone in Bengali.

This paper is organized as follows: section II presents few related works. Section III presents survey of language use. Section IV presents the root causes and the solutions to implement the accent understand ability. Section V describes the evaluation using two prototypes with users and findings. Section VI presents conclusion and future work.

## II. RELATED WORK

Researchers have given different ideas on various perspectives to design the user interface localization properly in mobile phone and other electronic devices. [6,7,8] state on issues such as text input, keystroke related analysis and space management where leaving about 30% extra space for characters from another language is preset. These papers have put important measure on localizing all parts of the user interface including help system and tool tips. These approaches provide grammatically correct complete or near-complete sentences and specific issues which are related to managing localization problems. Some researchers have recommended the way to identify best practices for localization.

Many technical concepts regarding text practice have been proposed which includes the use of Unicode text, to isolate text from code and to avoid text in bitmaps and icons. [9] presents a few cultural and political issues which involve the way to avoid the use of slangs, ethnocentric material or controversial maps in linguistics.

A few researchers have reported various techniques on the localization for mobile phones for Indian needs [10]; example includes the focus on localizing the predictive text input for Tamil mobile user interface. [11] presents the view of socio-cultural importance having localized user interfaces.

A recent literature [12] attempts to identify principles that one could use while translating and localizing the user interfaces, particularly the text used in commands and labels. Here they agree to standardize of the terms used in mobile phones to improve the usability and growth of mobile phone user interfaces.

### III. SURVEY OF LANGUAGE USAGE

I conducted a survey to see the usage of English and Indian languages, especially Bengali on mobile phones. I collected data about name, age, gender, education, mother tongue, duration of mobile phone use (year), current language on the phone, whether the user ever used a non-English language, languages that users speak at home, outside home, the languages in which they preferred to read newspapers, view news on TV and watch entertainment on TV. Table I presents the chart of language usage survey containing questions in left most column and the objects with numbers associated to it on the top most row of the chart.

TABLE I
CHART OF LANGUAGE USAGE SURVEY

| Questions | 01 | 02 | 03 | 04 | 05 |
|---|---|---|---|---|---|
| Name | | | | | |
| Age | | | | | |
| Gender | | | | | |
| Education | | | | | |
| Mother tongue | | | | | |
| Duration of mobile use (year) | | | | | |
| Current language on the phone | | | | | |
| Whether the user ever used a non-English language | | | | | |
| Users speak at home (language) | | | | | |
| Outside home (language) | | | | | |
| News channel language on TV | | | | | |
| Language of news paper | | | | | |
| Entertainment language on TV | | | | | |

I collected data from 25 users (11 females, 14 males) from Cooch Behar, a semi urban area situated in northern part of West Bengal. I ensured that I had few users in each of these education categories: Graduate, Post-graduate, 11th to 12th standards, 6th to 10th standards and 5th standard or less. Table II shows the average and standard deviation of age (objects) and their mobile phone usage. Table III gives the estimate of language usage in mobile phones. Table IV presents the number of English as the current language in each sample group based on education.

TABLE II
AGE AND MOBILE PHONE USAGE

| Metrics (year) | Average | Standard Deviation |
|---|---|---|
| Age (15 – 61) | 28.43 | 13.40 |
| Usage of Mobile | 4.73 | 2.61 |

TABLE III
PREFERENCE OF LANGUAGE USAGE

| Metrics (Language) | People Out of 25 | Percentage (%) |
|---|---|---|
| English | 23 | 92 |
| Bengali | 2 | 8 |

TABLE IV
ENGLISH AS CURRENT LANGUAGE IN MOBILE PHONES

| Metrics (education) | English/Number of Sample Group |
|---|---|
| Post-Graduate | 7/8 |
| Graduate | 5/5 |
| 11th – 12th | 5/5 |
| 6th – 10th | 4/5 |
| On or Below 5th | 2/2 |

When I asked whether any regional language (India) is available on the phone I found the data as given in the Table V below. 1 of 25 objects admitted that there was only English available in mobile phone. 15 among others reported that they had Bengali in their mobile phones.

TABLE V
NON ENGLISH LANGUAGE AVAILABLE IN MOBILE PHONE

| Metrics (regional language available) | Object Number |
|---|---|
| Nil | 1 |
| 1 | 9 |
| 2 | 8 |
| 3 | 5 |
| More than 3 | 2 |

In contrast with the languages used on the phone, people heavily preferred non-English Indian languages in other contexts. 25 out of 25 objects reported their mother tongue as Bengali, while all 25 reported speaking at home in Bengali. None of the users reported English as their mother tongue or as a language that they spoke at home. Only 5 of 25 objects reported English as one of the languages they speak outside home (including at work), of which 4 belonged to education group Post-graduate while 1 belonged to the education group 6th – 10th standard who is a student of 10th standard in an English medium school. None of the lower education groups reported using English either inside homes or outside.

TABLE VI
Use of News Paper and Its Language

| Metrics (News paper use at home) | Number of Objects | Language | | Education Group |
| | | Bengali | English | |
|---|---|---|---|---|
| Only 1 | 22 | 22 | 0 | All |
| More than 1 | 3 | 3 | 3 | Post-graduate |

TABLE VII
TV News and Its Language

| Metrics (TV news language) | Number of objects | Education Group |
|---|---|---|
| Bengali | 17 | All |
| Hindi | 3 | Graduate |
| English | 2 | Post-graduate |

TABLE VIII
Entertainment Language of TV

| Metrics (TV entertainment language) | Number of objects | Education Group |
|---|---|---|
| Bengali | 14 | All |
| Hindi | 8 | Graduate |
| English | 1[a] | 6th –10th |

a. 10th standard student from English medium school

Similarly, people preferred non-English Indian languages in other media including TV news, news paper and entertainment. Table 6 presents the language preference among the objects while for reading news in news paper. Table VII and VIII presents same but watching news and entertainment programmes on TV respectively.

Up to now it seems to be a primary resistor to use the phone in any Indian language especially in Bengali. I have found the root cause to be the usability issue arising out of accent understand ability of conventional terms used in day to day communication.

### IV. ROOT CAUSES AND CORRESPONDING TREATMENTS TO IMPLEMENT ACCENT UNDERSTAND ABILITY

I studied existing Bengali terms from the various mobile phones available in the market and shortlisted 250 frequently used terms. I then had practical talk with experts of Bengali to identify problems with current terms. I derived more conversational terms in the basis of natural accent understand ability. I pointed seven basic reasons and their solution to handle accent understand ability in Bengali. These are illustrated below.

*A. Few Bengali Terms are in Transliteration form of English*

I found many Bengali terms in transliterated form to its English one. Such a very use full term is Unlock. Currently it is same as **আনলক.** I have considered this as below.

| English term | Current Bengali term | Our Bengali term |
|---|---|---|
| Unlock | আনলক | লক খুলুন |

***Treatment:*** *Use less transliterated terms*

*B. Few Bengali Terms are in Very Much Formal Orientation to English Ones*

Many Bengali terms are in well form to their English. These kinds of term are troublesome to the people who do not talk in formal manner. One example is given below.

| English term | Current Bengali term | Our Bengali term |
|---|---|---|
| Option | বিকল্প | পছন্দ করুন |

***Treatment:*** *Use less formal terms*

*C. Wrong Translation*

I found that accent understand ability problem occur when wrong translation take place. The reason is the incorrect '*direct*' word translation without following the Bengali grammar. Example is as below.

| English term | Current Bengali term | Our Bengali term |
|---|---|---|
| Calling *Partha* | পার্থকে কল | পার্থকে কল করা হচ্ছে |

***Treatment:*** *Avoid grammatical mistake*

*D. Conventional Wrong Accent*

A few Bengali terms are uttered in general way, e.g., not following the grammar. The one is as below. Here the people utter **Missed Call** as **মিস্ কল** not as in English grammar. In general '*ed*' not is uttered by the people.

| English term | Current Bengali term | Our Bengali term |
|---|---|---|
| Missed Call | মিস্‌ড কল | মিস্ কল |

***Treatment:*** *Use more conventional accent*

*E. Wrong Translation of English Meaning in Bengali*

Often it is seen that few Bengali translations are wrong as per grammar. These terms must be avoided in order to avail easy understanding, hence increasing usability issues in mobile phones. One such example is shown below. Here meaning of **Reject** is misinterpreted as **বিদায়** but should be **বাতিল করুন**

| English term | Current Bengali term | Our Bengali term |
|---|---|---|
| Reject | বিদায় | বাতিল করুন |

***Treatment:*** *Avoid wrong translation of English and perform total translation but not by words*

*F. Use of Non Instructive Terms*

Unlike currently available Bengali strings in mobile phones, we have incorporated maximum portions of terms in highly interactive way. When user will look at the mobile, he/she will be comfortable to use the key terms, as in mobile is giving instructions tom perform the job. The same is illustrated in the following chart, where meaning of **বেক** is same as **Back.** Which is non interactive **আগের পৃষ্ঠায় ফিরে যান** instructs the user to go back to previous page.

| English term | Current Bengali term | Our Bengali term |
|---|---|---|
| Back | বেক | ফিরে যান |

***Treatment**: Use more instructive terms*

G. Use of Uncommon Terms

Sometimes uncommon terms which people do not use in daily work create problem in understanding. But নীরব রীতি is not commonly used to understand Mute, so I have derived চুপ করান Which is easily understandable.

| English term | Current Bengali term | Our Bengali term |
|---|---|---|
| Mute | নীরব রীতি | চুপ করান |

***Treatment**: Use commonly used terms*

## V. EVALUATION

I wanted to evaluate if the terms derived as per prescriptions presented above would lead to better usability in the prototype. It was not possible to evaluate all 250 terms in our experiment due to logistical constraints. I have shortlisted 39 well usable terms for the purpose of evaluation. These terms were based mainly upon 6 tasks such as, unlocking, making and receiving a call, contact related operations, and application based jobs, settings of phone and date and time setting. Table IX lists these 39 most frequently used terms, their current translations and recommendations using our prescriptions. Fig. 1 gives an over view of comparison between current and our terms. I also wanted to investigate the proper way of prototyping that might look forward for evaluating such a model product. I took textual terms for this purpose.

First, I made a 3-D 'thermacole' prototype with a printed image of a Nokia 2700 phone and cut out a window inside. I cut out similar size thermacole as of the mobile and glued to the behind the printed image. I printed various interface screens using our Bengali terms on a paper and inserted behind this window. In this prototype I could slide paper strips cut out of into a 'screen' (Fig. 2). The size of the printout was a slightly bigger to the actual mobile phone to enable easy operation.

The two sets of terms A and B from Table IX were evaluated with the help of this 3-D prototype with 6 users. I asked the users to perform the six tasks one after by using both sets of terms (sets A and B from Table 9). Half of the users used set B first and the rest used set A first. Their performance was measured in terms of time of completion to perform every task and number of errors they made. I counted an error when the user chose a wrong option than intended. At the end of the task, I asked users to give 'choice marks' to each set on a scale of 0-10. I also collected the qualitative feedback about inclination to use their phone in Bengali. Table X shows that the set B does better from set A in 3-D prototype by 69% and 22% in average number of errors and time of completion of tasks, respectively.

Though the 3-D prototype was good enough to handle by user, users were still unable to relate it with their actual mobile phones. At times, some users tried that task with their own mobile phones before performing it with the prototype. I tried to control the users performance to correlate in timely manner, but it was very hard in actual, because small time lag could give bad results.

TABLE IX
39 most commonly used terms on mobile phones, their current Bengali translations and my recommendations based on the given prescriptions suggested below

| Current English terms | Current Bengali translations (Set A) | Derived Bengali translations (Set B) |
|---|---|---|
| Unlock | আনলক | লক খুলুন |
| Keypad Active | কিপ্যাড আনলক করা হয়েছে | লক খুলেছে |
| Option | বিকল্প | পছন্দ করুন |
| Save | সেভ হল | সেভ করুন |
| Erase | ডিলিট | মুছুন |
| Messages | বার্তা গুলি | মেসেজ গুলি |
| Call | কল | ফোন করুন |
| Back | বেক | আগের পৃষ্ঠায় ফিরে যান |
| End | কল শেষ হল | কলটি সমাপ্ত |
| Cancel | বাতিল | কেন্সেল |
| Next | পরে | পরবর্তী পৃষ্ঠা |
| Create message | বার্তা সৃষ্টি করুন | মেসেজ লিখুন |
| Missed call | মিসড্ কল | মিস্ কল |
| Applications | প্রোগ্রাম গুলি | বিভিন্ন প্রয়োগ |
| Alarm | এলার্ম | সময় সংকেত |
| Music player | মিউজিক প্লেয়ার | সংগীত প্লেয়ার |
| Browser | ব্রাউজার | ব্রাউজ করুন |
| Calendar | কেলেন্ডার | দিন পঞ্জী |
| Time & Date | সময় ও তারিখ | দিন ও সময় |
| Multimedia | মাল্টিমিডিয়া | মিডিয়া গুলি |
| Menu | মেনু | তালিকা |
| View | দৃশ্য | সব দেখুন |
| Reject | বিদায় | বাতিল করুন |
| Calling *Partha* | পার্থকে কল | পার্থকে কল করা হছে |
| Call duration 2:12 | কলের সময় | শেষ কলের সময়কাল ২ মিনিট ১২ সেকেন্ড |
| *Partha* Calling | পার্থের কল | পার্থের কল এসেছে |
| Answer | গ্রহন | উত্তর দিন |
| Mute | নীরব রীতি | চুপ করান |
| Contacts | ফোন বুক | যোগাযোগ সমূহ |
| Log | কল লগ | কল নথি |
| Settings | সেটিংস | ব্যবস্থা |
| Details | বিবরণ | বিস্তৃত বিবরন |
| Add new contact | যোগাযোগ সৃষ্টি | নতুন নাম যোগ করুন |
| Edit contact | যোগাযোগ পরিবর্তন | নাম পরিবর্তন করুন |
| Search contact | যোগাযোগ খোঁজ | নামের খোঁজ করুন |
| Delete contact | যোগাযোগ ডিলিট | নাম বাতিল করুন |
| Game | গেম | খেলা |
| First name | প্রথম নাম | নাম |
| Last name | শেষ নাম | পদবী |

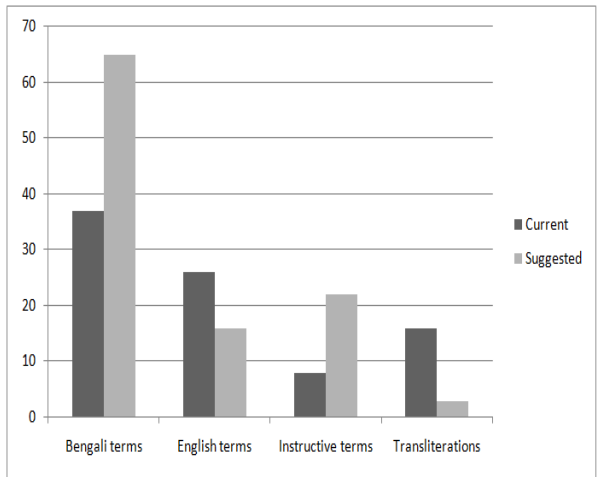

Fig. 1 Comparison between current and suggested Bengali terms

Therefore, I created two Microsoft Paint prototypes of Nokia 3230 on a HCL notebook and asked the users to perform the above said tasks (Fig. 3). One of the prototypes used the terms from set A, the other used terms from set B (from Table IX). I made sure that all the timed events in the displayed prototypes were as similar to real mobile phones as possible.

This time users could actually (almost) sensed a real mobile phone for performing tasks. Table X shows that set B did better over set A in errors and time. Overall, by combining results from all the two tests, I could see that set B did better than set A on errors and marks by 58% and 21% respectively.

TABLE X
Findings of usability tests, comparing terms in set A and set B using two different prototypes

| Metrics (prototype) | Average number of errors | | Time taken (sec.) | |
|---|---|---|---|---|
| | Set A | Set B | Set A | Set B |
| 3-D (N = 6) | 2.6 | 0.8 | 58 | 45 |
| MS Paint (N = 6) | 9.8 | 4.3 | 70 | 55 |
| Overall (N = 12) | 6.2 | 2.6 | 64 | 50 |

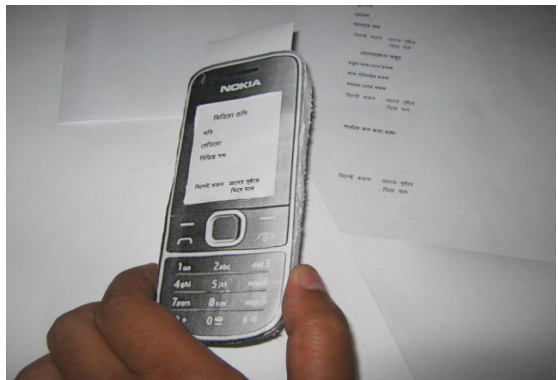

Fig. 2 Thermacole based prototype

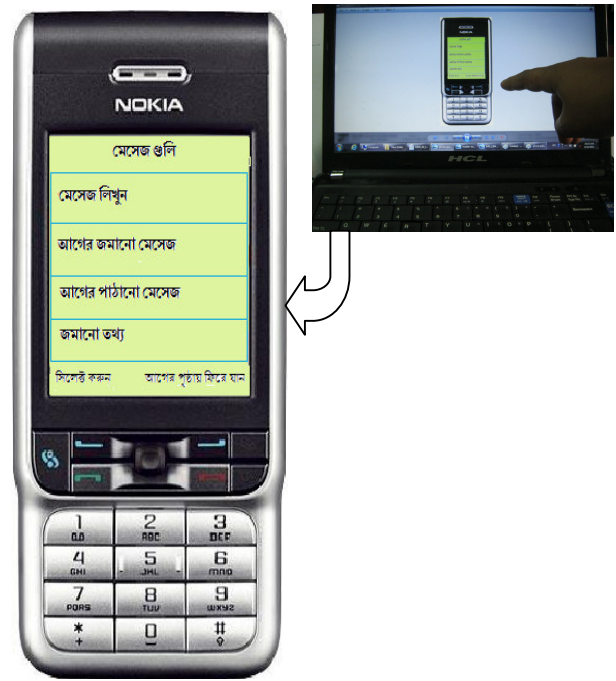

Fig. 3 Microsoft Paint based prototype displayed on notebook screen (right) and the larger view of the prototype (left)

VI. CONCLUSIONS

Our research presented a vivid view about language usage in West Bengal. The gist of our work includes that we (Indian people) prefer to speak in Indian languages where ever we are at home or outside, read language newspapers in regional or national language havoc, and watch entertainment and TV news in Indian languages with great interest. However, Indian people prefer to have their phones in English. As per prediction, highly educated people particularly use English on the phones. Regional or national Indian language usage on mobile phones is limited only with the people having lowest education.

In about 15% of the cases, phones do not support Non-English languages. This could happen in very old phones but also in the latest economical phones. Ignorance can be attributed to about 20% of people to English usage, because few people do not know whether their phones are enabled with any regional languages. I believe that one main reason for huge usage of English could be the pathetic usability of the Indian language interfaces, in particular poor term understand ability.

I acknowledge the shortcomings of our survey technique – I only asked few factual questions and did not seek for the causes why people prefer different language in different aspects on their mobile phones. Deep qualitative and informative interviews along with good ethnographic studies could explain the facts better and could be the tip of future investigations.

I identified seven types of causes with accent understand ability of Bengali terms and corresponding prescriptions for better and more usable accent understand ability. Depending on these criteria, I derived a bunch of 250 alternative Bengali terms for phone user interfaces that I believe would perform better. I evaluated 39 of the more commonly used terms and presented the findings in this paper. The other terms are also available but on request.

I found that less educated people performed fewer errors and preferred to use our terms in comparison with current terms. I also found that they are not satisfied with current terms and would prefer to use Bengali on their mobile phones if the terms are simpler and easier to understand.

In our experiments, I restricted myself to quantifiable parameters such as speed and errors. Future work using think-aloud techniques could investigate the reasons why some translations work better than others, probe the perceptions, thought processes and confusions caused by specific terms. This will enable further refinement of the prescriptions, helping them to become more useful.

Our study was restricted to a place of West Bengal and to Bengali user interfaces; I deeply feel that many other ways of linguistic based approaches could be applied in other extensively used regional languages in India.

Through-out the progress of our research work, I found the interesting effects of fidelity of the prototypes. I found the performance of users varied in terms of time and error, against the fidelity of prototypes. The investigation of language usage based on findings of significance could be a matter in future.

## REFERENCES

[1] Languages Spoken by More Than 10 Million People –MSN Encarta. http://encarta.msn.com/media_701500404/languages_spoken_by_more_than_10_million_people.html.
[2] List of languages by number of native speakers http://en.wikipedia.org/wiki/List_of_languages_by_number_of_native_speakers.
[3] State of Literacy, Census Report 2011, India.
[4] Annual report 2009-10, TRAI, Govt. of India.
[5] Joshi. A. Ganu, A. Chand, A. Parmar, V. Mathur, G. Keylekh(2004): a keyboard for Text entry in Indic Scripts CHI , April 24-29, 2004, 3-4.
[6] O'Reilly & Associates, Inc. Creating Applications With Mozilla.M. Young, The Technical Writer's Handbook. Mill Valley, 2009.
[7] Rohae, M. Keystroke-level analysis of Korean text entry methods on mobile phones. International Journal of Human-Computer Studies, Volume 60, Issues 5-6, Pages 545-563, 2004.
[8] Min, L. Chinese character entry for mobile phones: a longitudinal investigation. Interacting with Computers, Volume 17, Issue 2, Pages 121-146, 2005.
[9] Microsoft Inc. Best Practices for Globalization and Localization MSDN Visual Studio Develeoper Centre. http://msdn.microsoft.com/en-us/library/aa291552 (VS.71).aspx, 2009.
[10] Rajeshkannan, R, Nareshkumar, M., Ganesan, R. and Balakrishnan, R. Language Localization for Mobile Phones. http://tifac.velammal.org/CoMPC/articles/11.pdf , 2008.
[11] Marcus, A. Gould, E.W. Crosscurrents : Cultural Dimensions and Global Web User-Interface Design. Interactions, 2000, volume 7, issue 4, pages 32-46, 2000.
[12] N. Welankar, A. Joshi, K. Kanitkar, Principles for Simplifying Translation of Marathi Terms in Mobile Phones. In Proceedings of HCI2010/IDID2010.